\documentclass[prd, aps, superscriptaddress, preprintnumbers, twocolumn, floatfix, nofootinbib]{revtex4}

\usepackage{amsfonts}
\usepackage{amsmath}
\usepackage{amssymb}
\usepackage{bm}
\usepackage{dcolumn}
\usepackage{graphicx}   
\usepackage[latin1]{inputenc}
\usepackage{latexsym}
\usepackage{rotating}
\usepackage{hyperref}
\usepackage{graphicx}
\usepackage{color}

\usepackage{amsfonts}
\usepackage{amsmath}
\usepackage{amssymb}
\usepackage{bm}
\usepackage{dcolumn}
\usepackage{graphicx}
\usepackage[latin1]{inputenc}
\usepackage{latexsym}
\usepackage{rotating}
\usepackage{hyperref}

\newcommand\be{\begin{equation}}
\newcommand\ba{\begin{eqnarray}}
\newcommand\ee{\end{equation}}
\newcommand\ea{\end{eqnarray}}

\begin{document}

\title {Fast Radio Bursts from the Decay of Cosmic String Cusps}

\author{Robert Brandenberger}
\email{rhb@physics.mcgill.ca}
\affiliation{Physics Department, McGill University, Montreal, QC, H3A 2T8, Canada}

\author{Bryce Cyr}
\email{bryce.cyr@mail.mcgill.ca}
\affiliation{Physics Department, McGill University, Montreal, QC, H3A 2T8, Canada}

\author{Aditya Varna Iyer}
\email{aviyer@connect.ust.hk}
\affiliation{Department of Physics, The Hong Kong University of Science and Technology,
Clear Water Bay, Kowloon, Hong Kong, China, and \\
Physics Department, McGill University, Montreal, QC, H3A 2T8, Canada}

\date{\today}

\begin{abstract}

We explore the possibility that Fast Radio Bursts are due to the annihilation of cusps on
cosmic string loops. We compute the energy released in the annihilation events in
the radio region, the expected event rate, and the time scale of the bursts. We find
that the energy and event rates are sufficiently high and the time scale is sufficiently
small to explain the current data. We predict how the event rate will change as the
resolution of telescopes improves. Since the burst rate depends on the string
tension, future data will allow the determination of the tension.

\end{abstract}

\pacs{98.80.Cq}
\maketitle

\section{Introduction} 

In this Letter we explore the hypothesis that the observed Fast Radio Bursts (FRBs) 
could be due to the partial annihilation of cusps on cosmic string loops. Fast Radio
Bursts are bursts of electromagnetic radiation which have recently been observed
in large radio telescopes \cite{FRBobs}. They are characterized by a large energy flux (of the
order one {\it Jy}) over a very short time scale (of the order of several {\it ms})
and have been confirmed \cite{FRBlocal} as coming from cosmological
distances. Here we show that the partial annihilation of a cusp on a cosmic
string loop leads to a sufficiently large energy flux in radio wavelength over a 
sufficiently short time interval to be able to explain the FRB observations. We
compute the expected rate of FRB expected as a function of observation
threshold assuming that they are due to cosmic strings. We find a rate which
is sufficiently high to explain observations even if cusps only partially annihilate.

Cosmic strings \cite{VS, HK, RHBCSrev} are linear topological defect which exist as 
stable solutions of the equations of motion in many particle physics models beyond the 
{\it Standard Model}. If Nature is described by a model admitting string solutions, then a 
network of strings inevitably forms in the early universe and persists to the present
time \cite{Kibble}. Strings are lines of trapped energy and their gravitational effects
lead to signatures in cosmological observations. 

It can be shown that the network of cosmic strings approaches a {\it scaling solution}
in which the statistical properties of the string network are independent of time $t$ if
all lengths are scaled to $t$. The string scaling solution has two components: 
a random walk network of infinite strings with curvature radius of the same order
of magnitude as $t$, and a distribution of loops with number density 
distribution $n(R, t)$, where $R$ is the loop radius, and $n(R, t)$ is the number per
unit area per unit radius.

The distribution of strings (at least non-superconducting ones) is described by a single
free parameter, the string tension $\mu$, usually expressed in terms of the
dimensionless quantity $G \mu$, $G$ being Newton's gravitational constant (we
are working in natural units in which the speed of light and Planck's constant
are set to $1$). The strength of the cosmic string signals are proportional to 
$\mu = \eta^2$, where $\eta$ is the energy scale of the particle physics model
which gives the strings. Hence, searching for cosmological signatures of strings
is a way to probe particle physics beyond the {\it Standard Model} from ``top down'',
as opposed to accelerator searches which are more sensitive to lower values of
$\eta$ \cite{RHBrev1}. The current bound on the cosmic string tension from observations is
\cite{Dvorkin, PlanckCS}
\be \label{CSbound}
G \mu \, < \, 10^{-7} \, ,
\ee
(see also \cite{CSlimits}) which comes from precision measurements of the angular 
power spectrum of the cosmic microwave background (CMB). A tighter limit of \cite{GRbound}
\be
G \mu \, < \,  3 \times 10^{-9} \, ,
\ee
can be set from pulsar timing observations. They come from the amplitude
of the spectrum of gravitational waves emitted by oscillating string loops. But
since the loop distribution is less well established, this bound is less robust.

The long cosmic string segments yield interesting non-Gaussian signals in
cosmological observables such as CMB temperature maps \cite{KS, CMBTsearches},
CMB polarization maps \cite{Holder1} and 21cm redshift surveys \cite{Holder2, Oscar}.
Cosmic string loops were initially postulated to be the primary seeds of galaxies
\cite{CSearly}. This would require a string tension of the order of $G \mu \sim 10^{-6}$.
Such a large value of $G \mu$ is in conflict with CMB observations. Strings
with a value of $G \mu$ smaller than the bound given in (\ref{CSbound}) cannot
be the dominant mechanism of structure formation. However, since cosmic strings
form nonlinearities at arbitrarily large redshifts, they may play an interesting role
in structure formation. String loops may yield the seeds for high redshift super-massive
black holes \cite{SMBH} or for globular clusters \cite{GC}.

Cusps are a general feature of cosmic string loops. A cusp is a region where the
string doubles back on itself. In the approximation where the string is described
by a zero thickness line and hence described by the Nambu-Goto action, it can
be shown \cite{KibbleTurok} that any string loop develops at least one cusp per
oscillation time. However, strings which arise in gauge field theories have a
finite width (the width is the region within which the trapped potential energy
is confined). The string is a non-trivial configuration of gauge and scalar fields
(the gauge and scalar fields of the theory which yields the strings). A single
string is topologically stable. It is characterized by the fields winding the location
of the center of the string in a particular direction. An antistring is a field
configuration with the same energy profile, but with opposite winding. Locally, a cusp 
looks like a string-antistring pair and hence is completely unstable against decay into 
a burst of particles \cite{RHBcusp}. The decay products will be scalar and gauge
particles which decay into a jet of photons, neutrinos, and other stable particles
in a manner similar to how unstable particles decay and produce jets in 
particle accelerators. In the Nambu-Goto approximation the string velocity in
the center of mass frame of the string loop equals the speed of light. Hence,
the primary cusp decay particles will be highly beamed, and the secondary
decay products will also be beamed into an angle $\Theta$. The {\it cusp length}
$l_c$ can be defined as the length of the string where the two strands of the
string about the midpoint of the cusp have a separation closer than the string
width $w$ (which is of the order $\lambda^{-1/2}\eta^{-1}$ (where $\lambda$
is a dimensionless coupling constant). Possibly surprisingly, the cusp length
for a loop of radius $R$ is
\be \label{cusplength}
l_c \, \sim R^{2/3} w^{1/3} 
\ee
which is many orders of magnitude larger than the string width for string loops
present at late times. The fact that the cusp region is so long is responsible
for the fact that cusps may have important observational signatures \footnote{However,
it should be emphasized that the Nambu-Goto approximation completely breaks
down at the cusp, and that field theory back-reaction effects may greatly reduce
the effective cusp length.}.

Cusp decay has been considered a long time ago as a possible source of
ultra-high energy cosmic rays \cite{Jane1}, ultra-high energy neutrinos \cite{Jane2}
and gamma ray bursts \cite{gammaburst}. Cusp annihilation produces photons
with a continuous range of frequencies. In particular, photons in the radio range
will be produced. Hence, a cusp annihilation event will generate a burst in
radio wavelengths. Here we study whether the energy flux and event rate of
the cusp-induced bursts is sufficiently high to explain the data, and whether the
burst duration is sufficiently short. We find that these conditions are all
satisfied, and that hence string loop cusp annihilation provides a possible
explanation for the observed fast radio bursts.

\section{Analysis} 

A cosmic string cusp will decay into a jet whose primary particles are
quanta of the scalar and gauge fields which form the string. These, in
turn, decay into neutrinos, photons, and other stable particles. The
spectrum of decay particles is the same as seen in the decay of unstable
particles at accelerators. If $Q_f$ is the energy of the primary decay
particles, then the number $N(E)$ of photons received per unit area per unit energy
at a distance $d$ from the cusp is \cite{Jane2}
\be \label{numberdensity}
N(E) \, \sim \, \frac{1}{\Theta^2 d^2} \frac{\mu l_c}{Q_f^2} \bigl( \frac{Q_f}{E} \bigr)^{3/2}
\ee
for $E \ll Q_f$. It is reasonable to assume that the energy of the primary particles
is $Q_f \sim \eta$. The beaming angle $\Theta$ is given by
\be \label{beaming}
\Theta \, \sim \, \frac{1}{{\rm ln}(Q_f / {\rm GeV})} \, .
\ee

The energy flux $S$ of the observed FRBs is of the order $1 {\rm Jy}$, which in
natural units is
\be \label{limit}
S \, \sim \, 10^{-48} {\rm GeV}^3 \, ,
\ee
In order to explore the predicted detection rate for future telescopes which 
will have improved sensitivity we will multiply this flux by a factor $f$.

The first question we ask is to what distance $d(R)$ cusps on loops of radius
$R$ will lead to sufficient flux to be above the detection limit (\ref{limit}). 
Starting from the number density distribution (\ref{numberdensity}), multiplying
by the energy per photon $E$, integrating from $E = 0$ to $R$, 
inserting $Q_f = \eta$ and the expression (\ref{cusplength}) for the cusp
length, we obtain the following expression for the energy flux $S(E)$
\be \label{radioflux}
S(E, R, d) \, \sim \, \frac{1}{\Theta^2} \bigl( \frac{R}{d} \bigr)^2 \bigl( \frac{E}{\eta} \bigr)^{1/2}
\bigl( R \eta \bigr)^{-4/3} \eta^3 \, .
\ee
The energy $E$ which the radio telescopes are most sensitive to is
of the order $1 {\rm GHz} \sim 10^{-15} {\rm GeV}$. Making use of
\be
R \eta \, \sim \frac{R}{t_0} \bigl( t_0 m_{pl} \bigr) \frac{\eta}{m_{pl}} \, \sim \,
\frac{R}{t_0} 10^{60} (G \mu)^{1/2} \, ,
\ee
where $t_0$ is the present time and $m_{pl}$ is the Planck mass, the
condition
\be
S(E, R, d(R)) \, =  \, f S 
\ee
becomes
\be \label{limitrad}
d(R) \, \sim \, \frac{1}{\Theta} R^{1/3} t_0^{2/3} (G \mu)^{7/24} 10^3 f^{-1/2} \, .
\ee

The distance $d(R)$ must be compared to the mean separation $d_R$ of string loops of
radius $R$ (which is the expected distance of such a loop from us). This
distance is given by
\be
d_R^3 R n(R, t_0) \, = \, 1 \, ,
\ee
where $n(R, t)$ is the number density of string loops of radius $R$ (per unit radius)
at time $t$. According to the one scale model of the distribution of string loops
\cite{onescale}, the distribution of string loops at times $t \gg t_{eq}$ ($t_{eq}$ is
the time of equal matter and radiation) is given by
\be
n(R, t) \, = \, \alpha R^{-5/2} t_{eq}^{1/2} t^{-2} 
\ee
for
\be
\gamma G \mu t \, < \, R \, < \, t_{eq} \, ,
\ee
and by
\be
n(R, t) \, = \, \alpha R^{-2} t^{-2}
\ee
for
\be
t_{eq} \, < \, R \, < \, t \, .
\ee
Loops with radius $R < \gamma G \mu t$ decay in less than one Hubble
expansion time and can hence be neglected. In the above, $\gamma$ is
a constant which is determined by the strength of gravitational radiation from
a string loop. Numerical studies give $\gamma \sim 10$ \cite{CSgrav}.
The constant $\alpha$ is determined by cosmic string evolution simulations
and is expected to be of the order $1$.

Inserting all of the above formulas, the condition $d(R) > d_R$ becomes
\be
R^{1/2} \, < \, \frac{1}{\Theta^3} t_0^{1/2} (G \mu)^{21/24} f^{-3/2} 10^6 \,
\ee
and is easily satisfied for loops which dominate the string distribution
function and which also dominate the burst event rate, for all reasonable
values of $G \mu$.

Having shown that cosmic string cusp annihilations contain a sufficient
amount of energy to be seen by radio telescopes, we move on to
compute the expected rate of FRBs. This rate ${\cal R}$ is given by
\be
{\cal R} \, \sim \, \int_{\gamma G \mu t_0}^{t_0} dR d(R)^3 n(R, t_0) \frac{1}{R} P(R) \, ,
\ee
where the factor $1/R$ expresses the fact that there is of the order one cusp
for loop oscillation time (which is of the order $R$), and where $P(R)$
gives the probability that the jet from the cusp is beamed in direction of
the observer. This integral is dominated by the lower integration limit.
Inserting the expressions for $d(R)$ and for the string distribution, and taking
$P(R)$ to be a constant $P$ independent of $R$ we get
\be \label{result}
 {\cal R} \ \sim \, \frac{P}{\Theta^3} \alpha \gamma^{-3/2} (G \mu)_7^{-5/8} f^{-3/2} 10^{11} t_0^{-1} \, ,
\ee
where $(G \mu)_7$ is the value of $G \mu$ in units of $10^{-7}$.

The result (\ref{result}) shows that the predicted detection rate scales as $f^{-3/2}$ as
the detection limit is improved (i.e. $f$ decreases). The rate increases as $G \mu$
decreases since the increase in the number density of loops as $R$ decreases is a more
important effect than the decrease in the cusp energy. If we take $P = \Theta$
and use (\ref{beaming}) for the value of $\Theta$ we find a rate of bursts which
is sufficiently large to explain the current observations for any value of $G \mu$ below
the current bound.

We have now shown that the energy and event rate of cosmic string loop cusp
decays is sufficiently large to explain current observations. It remains to be
shown that the time scale of the cusp decay event is sufficiently small.
The time scale of a cusp event is given by
\be
t_{\rm cusp} \, \sim \, w^{2/3} R^{1/3} \, .
\ee
This can be argued in several ways. For example, we can argue that the
intrinsic time scale is enhanced by time dilation since the string cusp is
moving at a very high velocity in the observer's frame. Another way to
arrive at this time scale is to take the cusp configuration of \cite{RHBcusp}
and to ask for the time interval when the two strands of the string at
the cusp have moved apart by a distance more than the string width $w$
(see the Appendix). Using this equation we obtain
\be
T_{\rm cusp} \, \sim \, \bigl( \frac{R}{t_0} \bigr)^{1/3} (G \mu)_7^{1/3} 10^{-17} s \, ,
\ee
for the intrinsic time scale of the burst. The observed time scale of FRBs
is of the order $10^{-3} s$. The burst from a string loop will be spread out
over time by effects of propagation through the plasma. It is a consistency
check for our proposed explanation of FRBs that the initial time scale
is much smaller than the observed time.

\section{Optical Counterparts}

Bursts of photons produced by cosmic string cusp annihilations are not expected to 
be strictly localized to the radio regime. In fact, a prompt emission of radio waves 
should be accompanied by prompt emission of other frequencies, such as the 
optical photons. If the photon distribution follows the power law presented 
in equation (\ref{numberdensity}), the energy per unit area expected in optical frequencies is 
expected to be
\be
S_{\rm Opt}(E, R, d) \, \sim \, \frac{1}{\Theta^2 d^2} \frac{\mu l_c}{Q_f^2} \, ,
\int dE' E' \bigl( \frac{Q_f}{E'} \bigr)^{3/2}
\ee
where we perform the definite integral over the optical region, 
$1.8 \times 10^{-9}{\rm GeV} < E' < 3.1 \times 10^{-9}{\rm GeV}$ . We note that since 
$S(E,R,d)$ goes as $E^{1/2}$, the optical frequencies provide an enhanced 
energy flux over their radio counterpart. Comparing $S(E, R, d)$ for the two 
regimes allows us to quantify this enhancement. It is
\be
\eta_0 \, = \, \frac{S_{\rm Opt}(E,R,d)}{S_{\rm Rad}(E, R, d)} \, \sim \, 10^{3} \, ,
\ee
where $S_{Rad}(E, R, d)$ is the radio flux given in equation (\ref{radioflux}) 
evaluated at  $E = 1{\rm GHz}$ in natural units. Even though the signal is 
stronger (with this spectral distribution) in the optical, detection remains elusive for a 
couple of reasons. First of all, only one such FRB has been localized out of all the 
detections thus far. This presents the challenge of not necessarily knowing where 
to look for the optical burst from the majority of FRB signals. Secondly, optical 
telescopes typically possess large integration timescales, longer than 
one minute (see e.g. \cite{inttime, Lyutikov, Palomar}). Since the burst duration (at least in
the radio) is only on a millisecond timescale, optical bursts must be extremely 
bright to not get washed out over such long integration times.

It is possible to compute the observed magnitude $m$ of an optical burst, as would be 
seen here on earth, by a specific optical telescope. Such a formula has been derived 
in \cite{Lyutikov}
\be
m \, = \, 20.8 - 2.5 log_{10} \bigl( \frac{\eta_0 \tau_{ms} S_{Jy}}{T_{60}} \bigr) \, , 
\ee
where $S_{Jy}$ is the peak flux density of the radio burst in units of Janskys, $\tau_{ms}$ 
is the timescale of the burst in milliseconds, and $T_{60}$ is the integration time of the 
telescope in units of 60 seconds. For our model, generic FRB signals should give an 
optical burst magnitude of $m \sim 15$, which should be detectable by the Palomar 
Transient Factory (PTF) telescope \cite{Palomar}, as well as the LSST telescope in the future 
\cite{LSST}. It should be noted that searches have not yet detected 
an optical counterpart to the localized repeating FRB \cite{counterpart}.
A significant optical counterpart is currently a prediction of our model.

As has been known for some time \cite{gammaburst}, due to the $E^{-3/2}$ scaling of the number
density of photons in the cusp annihilation jet, cusp-induced high energy cosmic
ray bursts are expected to be far below the current detection threshold.  

\section{Conclusions and Discussion}

We have studied the emission of photons in the radio wavelength from 
cosmic string cusp annihilation events. We have shown that the
energy flux is sufficiently high at distances comparable to the
expected loop separation in order to be able to be seen by current
telescopes. We have computed the expected rate of such bursts
and shown that it is sufficiently large to explain current observations.
We have also computed the cusp decay time scale and found it
to be much smaller than the typical duration of an observed FRB
(the duration will in fact be dominated by propagation effects).
Hence, it appears that the cosmic string loop cusp decay provides
a mechanism for explaining the observed FRBs. Our calculation
predicts that the detection rate will increase as $f^{-3/2}$ as
the detection limit of the telescopes increase.

One FRB has been observed to be repeating \cite{Spitler}.
At the current level of understanding of cosmic string loop dynamics,
it is hard to address the question of whether bursts from cosmic
string cusp decay could be repeating. We know that there is at
least one cusp per loop per oscillation time. However, the
oscillation time for the loops which dominate our signal is of the
order $\gamma G \mu t_0$  and hence of the order of one year
if $G \mu = 10^{-11}$ (this value is chosen only for illustrative
purposes - it lies comfortably below the observational limits on
the string tension). There could, however, be many cusps on
a particular loop, and this could explain the repeating FRB, but
this explanation would also require each cusp to beam in the
same general direction. This discussion shows, however, that
we expect the properties of the string loop cusp annihilation process
regarding repeatability to be non-universal across the string loop
population.

Note that the mechanism which we are exploring is operative for all
types of cosmic strings. For superconducting strings \cite{Witten}
there is also direct emission of electromagnetic radiation \cite{Ostriker}.
This radiation is enhanced at string cusps, and this mechanism has
been explored as the origin or FRBs in some recent works \cite{Tanmay}
(see also \cite{oither}).
It has also recently been suggested that superconducting cosmic string 
loop collisions could lead to hot electromagnetic explosions which 
could explain FRBs \cite{Thompson}.

The major uncertainty concerning our mechanism is the effect of
back-reaction which could prevent the cusp from forming and developing
a length comparable to the one given in (\ref{cusplength}). 
For interesting analytical work on this issue see \cite{Olum1}, and 
for field theory simulations of cusps see \cite{Olum2} (see also \cite{Hindmarsh}
for a discussion of another code which can be used to study this
problem). We are leaving the discussion of back-reaction effects to future work.

The second uncertainty in our analysis is the assumption that the
$E^{-3/2}$ scaling of the number distribution of photons extends
down to radio frequencies. The $E^{-3/2}$ scaling is well established
down to the pion mass scale. Below that, we do not have any
direct measurements. However, it is well known from quantum
field theory that the number density diverges at least as fast
as $E^{-1}$ \cite{classic}.

\section*{Acknowledgement}
\noindent

We wish to thank Maxim Lyutikov, Ue-Li Pen and Pragya Chawla for
many discussions, and Brigitte Vachon, Andreas Warburton and Richard
Woodard for answering questions. The research at McGill is supported in
part by funds from NSERC and from the Canada Research Chair program.
AVI acknowledges support from a ``Lee Hysan Overseas Scholarship''
and a ``Paul and Mary Chu Overseas Summer Research
Travel Grant'' by the Department of Physics, HKUST.


\begin{thebibliography}{99}

\bibitem{FRBobs}
D.~R.~Lorimer, M.~Bailes, M.~A.~McLaughlin, D.~J.~Narkevic and F.~Crawford,
  ``A bright millisecond radio burst of extragalactic origin,''
  Science {\bf 318}, 777 (2007)
  doi:10.1126/science.1147532
  [arXiv:0709.4301 [astro-ph]];\\
  D.~Thornton {\it et al.},
  ``A Population of Fast Radio Bursts at Cosmological Distances,''
  Science {\bf 341}, no. 6141, 53 (2013)
  doi:10.1126/science.1236789
  [arXiv:1307.1628 [astro-ph.HE]].
  
\bibitem{FRBlocal}
S.~P.~Tendulkar {\it et al.},
  ``The Host Galaxy and Redshift of the Repeating Fast Radio Burst FRB 121102,''
  Astrophys.\ J.\  {\bf 834}, no. 2, L7 (2017)
  doi:10.3847/2041-8213/834/2/L7
  [arXiv:1701.01100 [astro-ph.HE]].
  
\bibitem{VS}
A. Vilenkin and E.P.S. Shellard, \textit{Cosmic Strings and other
Topological Defects} (Cambridge Univ. Press, Cambridge, 1994).

\bibitem{HK}
M.~B.~Hindmarsh and T.~W.~B.~Kibble, 
``Cosmic strings'', 
Rept.~Prog.~Phys.\ {\bf 58}, 477 (1995) 
[arXiv:hep-ph/9411342]. 

\bibitem{RHBCSrev}
R.~H.~Brandenberger,
  ``Topological defects and structure formation'',
  Int.\ J.\ Mod.\ Phys.\ A {\bf 9}, 2117 (1994)
  [arXiv:astro-ph/9310041].

\bibitem{Kibble}
T.~W.~B.~Kibble,
  ``Phase Transitions In The Early Universe'',
  Acta Phys.\ Polon.\  B {\bf 13}, 723 (1982);\\
  T.~W.~B.~Kibble,
  ``Some Implications Of A Cosmological Phase Transition'',
  Phys.\ Rept.\  {\bf 67}, 183 (1980).
  
 \bibitem{RHBrev1}
R.~H.~Brandenberger,
  ``Probing Particle Physics from Top Down with Cosmic Strings'',
  Universe {\bf 1}, no. 4, 6 (2013)
  [arXiv:1401.4619 [astro-ph.CO]].
  
  \bibitem{Dvorkin}
T.~Charnock, A.~Avgoustidis, E.~J.~Copeland and A.~Moss,
  ``CMB Constraints on Cosmic Strings and Superstrings'',
  arXiv:1603.01275 [astro-ph.CO];\\
C.~Dvorkin, M.~Wyman and W.~Hu,
  ``Cosmic String constraints from WMAP and the South Pole Telescope'',
  Phys.\ Rev.\ D {\bf 84}, 123519 (2011)
  [arXiv:1109.4947 [astro-ph.CO]].
 
\bibitem{PlanckCS}
   P.~A.~R.~Ade {\it et al.}  [Planck Collaboration],
  ``Planck 2013 results. XXV. Searches for cosmic strings and other topological defects'',
  Astron.\ Astrophys.\  {\bf 571}, A25 (2014)
  [arXiv:1303.5085 [astro-ph.CO]].
 
\bibitem{CSlimits}
L.~Pogosian, S.~H.~H.~Tye, I.~Wasserman and M.~Wyman,
  ``Observational constraints on cosmic string production during brane
  inflation'',
  Phys.\ Rev.\  D {\bf 68}, 023506 (2003)
  [Erratum-ibid.\  D {\bf 73}, 089904 (2006)]
  [arXiv:hep-th/0304188];\\
M.~Wyman, L.~Pogosian and I.~Wasserman,
  ``Bounds on cosmic strings from WMAP and SDSS'',
  Phys.\ Rev.\  D {\bf 72}, 023513 (2005)
  [Erratum-ibid.\  D {\bf 73}, 089905 (2006)]
  [arXiv:astro-ph/0503364];\\
A.~A.~Fraisse,
  ``Limits on Defects Formation and Hybrid Inflationary Models with
  Three-Year WMAP Observations'',
  JCAP {\bf 0703}, 008 (2007)
  [arXiv:astro-ph/0603589];\\
U.~Seljak, A.~Slosar and P.~McDonald,
  ``Cosmological parameters from combining the Lyman-alpha forest with CMB,
  galaxy clustering and SN constraints'',
  JCAP {\bf 0610}, 014 (2006)
  [arXiv:astro-ph/0604335];\\
  R.~A.~Battye, B.~Garbrecht and A.~Moss,
  ``Constraints on supersymmetric models of hybrid inflation'',
  JCAP {\bf 0609}, 007 (2006)
  [arXiv:astro-ph/0607339];\\
R.~A.~Battye, B.~Garbrecht, A.~Moss and H.~Stoica,
  ``Constraints on Brane Inflation and Cosmic Strings'',
  JCAP {\bf 0801}, 020 (2008)
  [arXiv:0710.1541 [astro-ph]];\\
N.~Bevis, M.~Hindmarsh, M.~Kunz and J.~Urrestilla,
  ``CMB power spectrum contribution from cosmic strings using  field-evolution
  simulations of the Abelian Higgs model'',
  Phys.\ Rev.\  D {\bf 75}, 065015 (2007)
  [arXiv:astro-ph/0605018];\\
N.~Bevis, M.~Hindmarsh, M.~Kunz and J.~Urrestilla,
  ``Fitting CMB data with cosmic strings and inflation'',
  Phys.\ Rev.\ Lett.\  {\bf 100}, 021301 (2008)
  [astro-ph/0702223 [ASTRO-PH]];\\
R.~Battye and A.~Moss,
  ``Updated constraints on the cosmic string tension'',
 Phys.\ Rev.\ D {\bf 82}, 023521 (2010)
  [arXiv:1005.0479 [astro-ph.CO]].

\bibitem{GRbound}
J.~J.~Blanco-Pillado, K.~D.~Olum and B.~Shlaer,
  ``The number of cosmic string loops,''
  Phys.\ Rev.\ D {\bf 89}, no. 2, 023512 (2014)
  doi:10.1103/PhysRevD.89.023512
  [arXiv:1309.6637 [astro-ph.CO]].
  
\bibitem{KS}
N.~Kaiser and A.~Stebbins,
  ``Microwave Anisotropy Due To Cosmic Strings'',
  Nature {\bf 310}, 391 (1984).

\bibitem{CMBTsearches}
R.~J.~Danos and R.~H.~Brandenberger,
  ``Canny Algorithm, Cosmic Strings and the Cosmic Microwave Background'',
  Int.\ J.\ Mod.\ Phys.\ D {\bf 19}, 183 (2010)
  [arXiv:0811.2004 [astro-ph]];\\
S.~Amsel, J.~Berger and R.~H.~Brandenberger,
  ``Detecting Cosmic Strings in the CMB with the Canny Algorithm'',
  JCAP {\bf 0804}, 015 (2008)
  [arXiv:0709.0982 [astro-ph]];\\
A.~Stewart and R.~Brandenberger,
  ``Edge Detection, Cosmic Strings and the South Pole Telescope'',
 JCAP {\bf 0902}, 009 (2009)
  [arXiv:0809.0865 [astro-ph]];\\
L.~Hergt, A.~Amara, R.~Brandenberger, T.~Kacprzak and A.~Refregier,
  ``Searching for Cosmic Strings in CMB Anisotropy Maps using Wavelets and Curvelets,''
  arXiv:1608.00004 [astro-ph.CO];\\
J.~D.~McEwen, S.~M.~Feeney, H.~V.~Peiris, Y.~Wiaux, C.~Ringeval and F.~R.~Bouchet,
  ``Wavelet-Bayesian inference of cosmic strings embedded in the cosmic microwave background,''
  arXiv:1611.10347 [astro-ph.IM];\\
 R.~Ciuca and O.~F.~Hernández,
  ``A Bayesian Framework for Cosmic String Searches in CMB Maps,''
  arXiv:1706.04131 [astro-ph.CO].
  
\bibitem{Holder1}
R.~J.~Danos, R.~H.~Brandenberger and G.~Holder,
  ``A Signature of Cosmic Strings Wakes in the CMB Polarization'',
  Phys.\ Rev.\ D {\bf 82}, 023513 (2010)
  [arXiv:1003.0905 [astro-ph.CO]].
  
\bibitem{Holder2}
R.~H.~Brandenberger, R.~J.~Danos, O.~F.~Hernandez and G.~P.~Holder,
  ``The 21 cm Signature of Cosmic String Wakes'',
  JCAP {\bf 1012}, 028 (2010)
  [arXiv:1006.2514 [astro-ph.CO]].
 
\bibitem{Oscar}
O.~F.~Hernandez,
  ``Wouthuysen-Field absorption trough in cosmic string wakes,''
  Phys.\ Rev.\ D {\bf 90}, no. 12, 123504 (2014)
  doi:10.1103/PhysRevD.90.123504
  [arXiv:1403.7522 [astro-ph.CO]].

\bibitem{CSearly}
A.~Vilenkin,
  ``Cosmological Density Fluctuations Produced by Vacuum Strings'',
  Phys.\ Rev.\ Lett.\  {\bf 46}, 1169 (1981)
  Erratum: [Phys.\ Rev.\ Lett.\  {\bf 46}, 1496 (1981)].
  doi:10.1103/PhysRevLett.46.1169, 10.1103/PhysRevLett.46.1496; \\
N.~Turok and R.~H.~Brandenberger,
  ``Cosmic Strings And The Formation Of Galaxies And Clusters Of Galaxies'',
  Phys.\ Rev.\ D {\bf 33}, 2175 (1986);\\
H. Sato, ``Galaxy Formation by Cosmic Strings'',
  Prog. Theor. Phys.\  {\bf 75}, 1342 (1986);\\
A. Stebbins, ``Cosmic Strings and Cold Matter'',
  Ap. J. (Lett.) {\bf 303}, L21 (1986).
  
\bibitem{SMBH}
S.~F.~Bramberger, R.~H.~Brandenberger, P.~Jreidini and J.~Quintin,
  ``Cosmic String Loops as the Seeds of Super-Massive Black Holes,''
  JCAP {\bf 1506}, no. 06, 007 (2015)
  doi:10.1088/1475-7516/2015/06/007
  [arXiv:1503.02317 [astro-ph.CO]].
  
\bibitem{GC}
A.~Barton, R.~H.~Brandenberger and L.~Lin,
  ``Cosmic Strings and the Origin of Globular Clusters,''
  JCAP {\bf 1506}, no. 06, 022 (2015)
  doi:10.1088/1475-7516/2015/06/022
  [arXiv:1502.07301 [astro-ph.CO]];\\
  L.~Lin, S.~Yamanouchi and R.~Brandenberger,
  ``Effects of Cosmic String Velocities and the Origin of Globular Clusters,''
  JCAP {\bf 1512}, no. 12, 004 (2015)
  doi:10.1088/1475-7516/2015/12/004
  [arXiv:1508.02784 [astro-ph.CO]].
  
\bibitem{KibbleTurok}
T.~W.~B.~Kibble and N.~Turok,
  ``Selfintersection of Cosmic Strings,''
  Phys.\ Lett.\  {\bf 116B}, 141 (1982).
  doi:10.1016/0370-2693(82)90993-5
  
\bibitem{RHBcusp}
 R.~H.~Brandenberger,
  ``On the Decay of Cosmic String Loops,''
  Nucl.\ Phys.\ B {\bf 293}, 812 (1987).
  doi:10.1016/0550-3213(87)90092-7
      
\bibitem{Jane1}
J.~H.~MacGibbon and R.~H.~Brandenberger,
  ``Gamma-ray signatures from ordinary cosmic strings,''
  Phys.\ Rev.\ D {\bf 47}, 2283 (1993)
  doi:10.1103/PhysRevD.47.2283
  [astro-ph/9206003];\\
U.~F.~Wichoski, J.~H.~MacGibbon and R.~H.~Brandenberger,
  ``High-energy neutrinos, photons and cosmic ray fluxes from VHS cosmic strings,''
  Phys.\ Rev.\ D {\bf 65}, 063005 (2002)
  doi:10.1103/PhysRevD.65.063005
  [hep-ph/9805419].
  
\bibitem{Jane2}
J.~H.~MacGibbon and R.~H.~Brandenberger,
  ``High-energy neutrino flux from ordinary cosmic strings,''
  Nucl.\ Phys.\ B {\bf 331}, 153 (1990).
  doi:10.1016/0550-3213(90)90020-E
 
\bibitem{gammaburst}
R.~H.~Brandenberger, A.~T.~Sornborger and M.~Trodden,
  ``Gamma-ray bursts from ordinary cosmic strings,''
  Phys.\ Rev.\ D {\bf 48}, 940 (1993)
  doi:10.1103/PhysRevD.48.940
  [hep-ph/9302254].

\bibitem{onescale}
A.~Vilenkin,
  ``Cosmic Strings,''
    Phys.\ Rev.\ D {\bf 24}, 2082 (1981);\\
  T.~W.~B.~Kibble,
  ``Evolution of a system of cosmic strings,''
  Nucl.\ Phys.\ B {\bf 252}, 227 (1985)
  [Nucl.\ Phys.\ B {\bf 261}, 750 (1985)].
  
 \bibitem{CSgrav}
  T.~Vachaspati and A.~Vilenkin,
  ``Gravitational Radiation from Cosmic Strings,''
  Phys.\ Rev.\ D {\bf 31}, 3052 (1985).
  
 \bibitem{inttime}
 N.~M.~Law {\it et al.},
  ``Evryscope science: exploring the potential of all-sky gigapixel-scale telescopes,''
  Publ.\ Astron.\ Soc.\ Pac.\  {\bf 127}, 234 (2015)
  doi:10.1086/680521
  [arXiv:1501.03162 [astro-ph.IM]];\\
 S.~Griffin, D.~Hanna and A.~Gilbert,
  ``Searching for Fast Optical Transients using VERITAS Cherenkov Telescopes,''
  IAU Symp.\  {\bf 285}, 321 (2012)
  doi:10.1017/S1743921312000932
  [arXiv:1206.6535 [astro-ph.HE]].
  
 \bibitem{Lyutikov}
 M.~Lyutikov and D.~R.~Lorimer,
  ``How else can we detect Fast Radio Bursts?,''
  Astrophys.\ J.\  {\bf 824}, no. 2, L18 (2016)
  doi:10.3847/2041-8205/824/2/L18
  [arXiv:1605.01468 [astro-ph.HE]].
  
 \bibitem{Palomar}
 N.~M.~Law {\it et al.},
  ``The Palomar Transient Factory: System Overview, Performance and First Results,''
  Publ.\ Astron.\ Soc.\ Pac.\  {\bf 121}, 1395 (2009)
  doi:10.1086/648598
  [arXiv:0906.5350 [astro-ph.IM]].
  
 \bibitem{LSST}
 P.~A.~Abell {\it et al.} [LSST Science and LSST Project Collaborations],
  ``LSST Science Book, Version 2.0,''
  arXiv:0912.0201 [astro-ph.IM].
  
 \bibitem{counterpart}
 S.~Q.~Xi, P.~H.~T.~Tam, F.~K.~Peng and X.~Y.~Wang,
  ``Search for GeV counterparts to fast radio bursts with Fermi,''
  Astrophys.\ J.\  {\bf 842}, no. 1, L8 (2017)
  doi:10.3847/2041-8213/aa74cf
  [arXiv:1705.03657 [astro-ph.HE]];\\
  P.~Scholz {\it et al.},
  ``Simultaneous X-ray, gamma-ray, and Radio Observations of the repeating Fast Radio Burst FRB 121102,''
  arXiv:1705.07824 [astro-ph.HE].
 
 \bibitem{Spitler}
L.~G.~Spitler {\it et al.},
  ``A Repeating Fast Radio Burst,''
  Nature {\bf 531}, 202 (2016)
  doi:10.1038/nature17168
  [arXiv:1603.00581 [astro-ph.HE]];\\
  P.~Scholz {\it et al.},
  ``The repeating Fast Radio Burst FRB 121102: Multi-wavelength observations and additional bursts,''
  Astrophys.\ J.\  {\bf 833}, no. 2, 177 (2016)
  doi:10.3847/1538-4357/833/2/177
  [arXiv:1603.08880 [astro-ph.HE]].
  
 \bibitem{Witten}
 E.~Witten,
  ``Superconducting Strings,''
  Nucl.\ Phys.\ B {\bf 249}, 557 (1985).
  doi:10.1016/0550-3213(85)90022-7
  
 \bibitem{Ostriker}
 J.~P.~Ostriker, A.~C.~Thompson and E.~Witten,
  ``Cosmological Effects of Superconducting Strings,''
  Phys.\ Lett.\ B {\bf 180}, 231 (1986).
  doi:10.1016/0370-2693(86)90301-1
  
 \bibitem{Tanmay}
 T.~Vachaspati,
  ``Cosmic Sparks from Superconducting Strings,''
  Phys.\ Rev.\ Lett.\  {\bf 101}, 141301 (2008)
  doi:10.1103/PhysRevLett.101.141301
  [arXiv:0802.0711 [astro-ph]];\\
  L.~V.~Zadorozhna and B.~I.~Hnatyk,
  ``Electromagnetic emission bursts from the near-cusp regions of superconducting cosmic strings,''
  Ukr.\ J.\ Phys.\  {\bf 54}, 1149 (2009);\\
 Y.~F.~Cai, E.~Sabancilar and T.~Vachaspati,
  ``Radio bursts from superconducting strings,''
  Phys.\ Rev.\ D {\bf 85}, 023530 (2012)
  doi:10.1103/PhysRevD.85.023530
  [arXiv:1110.1631 [astro-ph.CO]];\\
 Y.~F.~Cai, E.~Sabancilar, D.~A.~Steer and T.~Vachaspati,
  ``Radio Broadcasts from Superconducting Strings,''
  Phys.\ Rev.\ D {\bf 86}, 043521 (2012)
  doi:10.1103/PhysRevD.86.043521
  [arXiv:1205.3170 [astro-ph.CO]];\\
  L.~V.~Zadorozhna,
  ``Fast radio bursts as electromagnetic radiation fro cusps on 
  superconducting cosmic strings'',
  Adv. in Astronomy and Space Physics {\bf 5}, 43 (2015);\\
  J.~Ye, K.~Wang and Y.~F.~Cai,
  ``Superconducting Cosmic Strings as Sources of Cosmological Fast Radio Bursts,''
  arXiv:1705.10956 [astro-ph.HE].

\bibitem{other} 
V.~Berezinsky, B.~Hnatyk and A.~Vilenkin,
  ``Gamma-ray bursts from superconducting cosmic strings,''
  Phys.\ Rev.\ D {\bf 64}, 043004 (2001)
  doi:10.1103/PhysRevD.64.043004
  [astro-ph/0102366];\\
  A.~Gruzinov and A.~Vilenkin,
  ``Fireballs from Superconducting Cosmic Strings,''
  JCAP {\bf 1701}, no. 01, 029 (2017)
  doi:10.1088/1475-7516/2017/01/029
  [arXiv:1608.05396 [astro-ph.HE]].
  
\bibitem{Thompson}
C.~Thompson,
  ``Tiny Electromagnetic Explosions,''
  arXiv:1703.00393 [astro-ph.HE];\\
  C.~Thompson,
  ``Giant Primeval Magnetic Dipoles,''
  arXiv:1703.00394 [astro-ph.HE].
 
\bibitem{Olum1}
 J.~J.~Blanco-Pillado and K.~D.~Olum,
  ``The Form of cosmic string cusps,''
  Phys.\ Rev.\ D {\bf 59}, 063508 (1999)
  doi:10.1103/PhysRevD.59.063508
  [gr-qc/9810005].
  
\bibitem{Olum2} 
K.~D.~Olum and J.~J.~Blanco-Pillado,
  ``Field theory simulation of Abelian Higgs cosmic string cusps,''
  Phys.\ Rev.\ D {\bf 60}, 023503 (1999)
  doi:10.1103/PhysRevD.60.023503
  [gr-qc/9812040].
  
 \bibitem{Hindmarsh}
 M.~Hindmarsh, J.~Lizarraga, J.~Urrestilla, D.~Daverio and M.~Kunz,
  ``Scaling from gauge and scalar radiation in Abelian Higgs string networks,''
  arXiv:1703.06696 [astro-ph.CO].
 
\bibitem{classic} 
F.~Bloch and A.~Nordsieck,
  ``Note on the Radiation Field of the electron,''
  Phys.\ Rev.\  {\bf 52}, 54 (1937).
  doi:10.1103/PhysRev.52.54;\\
  A.~Nordsieck,
  ``The Low Frequency Radiation of a Scattered Electron,''
  Phys.\ Rev.\  {\bf 52}, 59 (1937).
  doi:10.1103/PhysRev.52.59
  
\end{thebibliography}
\end{document}